\documentstyle[aps,prl,epsf,floats]{revtex}

\begin{document}

\ifpreprintsty \else
\twocolumn[\hsize\textwidth\columnwidth\hsize\csname@twocolumnfalse%
\endcsname \fi
 
\draft
\title{Anomalous pinning behavior \\in an incommensurate two-chain model of friction}
\author{Takaaki Kawaguchi}
\address{Department of Technology, Faculty of Education, Shimane University, 
1060 Nishikawatsu, Matsue 690-8504, Japan}
\author{Hiroshi Matsukawa}
\address{Department of Physics, Osaka University, 
1-16 Machikaneyama Toyonaka, Osaka 560-0043, Japan}
\date{\today}
\maketitle
\begin{abstract}
Pinning phenomena in an incommensurate two-chain model of friction are studied numerically.
The pinning effect due to the breaking of analyticity exists in the present model. 
The pinning behavior is, however, quite different from 
that for the breaking of the analyticity state of the Frenkel-Kontorova model. 
When the elasticity of chains or the strength of interchain interaction is changed, 
pinning force and maximum static frictional force show anomalously complicated behavior 
accompanied by a successive phase transition 
and they vanish completely under certain conditions. 
\vspace*{0.7cm}

\end{abstract}
\ifpreprintsty \else
] \fi              

\narrowtext
%
\section{Introduction}

In recent years, the study of friction has been attracting much attention 
in physics \cite{sliding}.
Nano-scale frictional phenomena have been examined experimentally 
using frictional force microscopes \cite{mate}, 
quartz micro-valance techniques \cite{krim} and so on.
In theoretical studies, 
the Frenkel-Kontorova (FK) model \cite{fk} and its related ones have been employed 
as a promising model of 
such nano-scale friction by several researchers \cite{soko1,shinjo,gra,elmer}.
The FK model, in general, consists of an atomic chain 
on a substrate with periodic potential. 
In the chain harmonic force works between neighboring atoms.
When the mean atomic distance and the period of the potential is incommensurate, 
the FK model shows a phase transition, 
which has been discussed in detail by Aubry and coworkers \cite{aubry1,peyrard}.
Hence this phase transition is called the Aubry transition.
The Aubry transition has following features.
When the amplitude of the substrate potential is smaller than a certain critical value, 
the lowest phonon excitation is gapless and therefore a free sliding mode appears.
This means vanishing maximum static frictional force.
Above the critical amplitude, however, 
the atoms in the chain are pinned strongly nearby the potential minima and 
a finite gap exists in the phonon excitation. 
This state is called the breaking of the analyticity state. 
Then finite energy is needed to slide the chain, 
and therefore the maximum static frictional force becomes finite.
The extended FK model which consists of interacting two deformable 
chains also have been investigated so far.
The static structural properties of two-chain models have been investigated 
in Refs. \cite{theod,cout,allro,ishi}, 
where each chain is often treated as a continuum elastic line. 
The continuum approximation works effectively in the study on 
the commensurate-incommensurate transition \cite{cout}.
However, the pinning effect, that arises from the discrete nature of lattices, 
are smeared out inevitably.
In other words, the Aubry transition never occurs in the continuum models.
On the basis of a two-chain model with discrete lattice structures, 
Matsukawa and Fukuyama investigated the static and the kinetic frictional forces \cite{matsu1,matsu2}. 
The model proposed in their study consists of two atomic chains, where interchain atomic 
force works between atoms in one chain and in another and 
harmonic force works between neighbor atoms in each chain. 
In some cases of elastic parameters of chains, 
they found that the maximum static frictional force for the two-chain model becomes larger than 
that for the FK model
with the same strength of the interchain interaction.
Furthermore, they discussed the relationship between the strength of 
the maximum static frictional force and 
the velocity dependence of kinetic frictional force.

In this paper, we revisit 
the two-chain model of friction employed in Ref. \cite{matsu1} 
and examine the frictional phenomena in a wide range of model parameters. 
In particular, pinned states are investigated thoroughly 
in connection with the breaking of analyticity state due to the Aubry transition. 
It turns out that 
the maximum static frictional force shows complicated behavior 
against the change in elastic parameters and vanishes completely in certain conditions.
This anomalous pinning behavior is discussed in relation to the static lattice 
structures.
We also focus on the velocity dependence of the kinetic frictional force 
in sliding states.
\section{Two-chain model of friction}

The two-chain model of friction employed here is summarized in the following \cite{matsu1}.
We consider two atomic chains, i.e., an upper chain and a lower chain.
Each atom has one-dimensional degree of freedom parallel to the chain.
Intrachain interaction with harmonic form 
and interchain interaction 
are taken into consideration.
The effects of energy dissipation are assumed to be 
proportional to the difference between the velocity of each atom 
and that of the center of gravity
of the chain.
The upper chain is driven by the external force parallel to the
chain.
Assuming overdamped motion, we get the equations of
motion of the atoms in the upper and the lower chains given by
\begin{eqnarray}
m_a\gamma_a(\dot{u}_i-\langle\dot{u}_i\rangle_i)
&=&  K_a(u_{i+1}+u_{i-1}-2u_i) \nonumber \\
&& +\sum^{N_b}_{j\in b} F_I(u_i-v_j) +F_{ex},\\
m_b\gamma_b(\dot{v}_i-\langle\dot{v}_i\rangle_i)
&=&  K_b(v_{i+1}+v_{i-1}-2v_i) \nonumber \\
&& +\sum^{N_a}_{j\in a} F_I(v_i-u_j) - K_s(v_i-ic_b),
\end{eqnarray}
where $u_i$ ($v_i$), $m_a$ ($m_b$), $\gamma_a$ ($\gamma_b$), 
$K_a$ ($K_b$), and $N_a$ ($N_b$) 
are the position of the $i$th atom, 
the atomic mass, 
the parameter of energy dissipation, 
the strength of the interatomic force, 
and the number of atoms in the upper (lower) chain, respectively.
$K_s$ denotes 
the strength of the interatomic force between the lower chain and 
the substrate, which is necessary to bind the lower chain.
$\langle\;\;\rangle_i$ 
represents the average with respect to $i$.
$F_I$ and $F_{ex}$ are the interchain force between the two atomic
chains and the external force, respectively.
The interchain atomic potential is chosen as follows:
\begin{equation}
U_{I}=-\frac{K_I}{2}\exp{\left[-4\left(\frac{x}{c_b}\right)^2\right]},
\end{equation}
where $K_I$ is the strength of the interchain potential, and $c_b$ the
mean atomic spacing of the lower chain.
The interatomic force is given by $F_I(x)=-\frac{d}{dx}U_{I}$.
The time-averaged total frictional force of the present model 
is given by 
\begin{equation}
F^{\rm fric}= 
- \sum^{N_a}_{i\in a}\sum^{N_b}_{j\in b}\left\langle F_I( v_i-u_j)
\right\rangle_{t} = N_a \left\langle F_{ex} \right\rangle_{t}.
\end{equation}
This expression of the frictional force is
valid for both static and kinetic ones.
The form of interchain force, $F_I(x)=-\frac{d}{dx}{U_{I}}$, 
is approximated in the following two limiting cases \cite{kawa1}.
When the atoms in the lower chain are rigid and fixed at the regular sites, 
which corresponds to the case of
$K_b/K_a \rightarrow \infty$ or $K_s/K_a \rightarrow \infty$, 
the interchain force that acts on the upper chain is approximated by 
one term in a Fourier series: 
\begin{eqnarray}
\left. \sum^{N_b}_{j\in b} F_I(u_i-v_j) \right|_{v_j=jc_b}
&\simeq&
-0.47{K_I}\sin\left( 2\pi\frac{u_i}{c_b} \right) .
\end{eqnarray}
Then the two-chain model is reduced to the FK model.
In the opposite limit, when the upper chain is fixed at the regular sites 
($K_a/K_b\gg 1$), 
the interchain interaction that acts on the lower chain is
also described by one term in a Fourier series: 
\begin{eqnarray}
\left. \sum^{N_a}_{i\in a} F_I(u_i-v_j) \right|_{u_i=ic_a}
&\simeq& - 0.83{K_I}\sin\left(2\pi\frac{v_j}{c_a} \right),
\end{eqnarray}
where $c_a$ is the mean atomic spacing of the upper chain.
These approximations on the interchain force 
are valid only in the above two limiting cases on elastic parameters, 
and they may break in intermediate cases. 
This is a crucial point in the following discussion.
\section{Numerical method}

For the numerical simulation of the model, the Runge-Kutta (RK) formula is employed to solve 
the equations of motion.
The periodic boundary conditions are employed in both chains. 
Hence, the ratio $c_a/c_b$ is equal to $N_b/N_a$, where $N_a$ and $N_b$ are 
the numbers of atoms in the upper and lower bodies.
\begin{equation}
\frac{c_a}{c_b}=\frac{N_b}{N_a}=\frac{233}{144}=1.618\cdots,
\end{equation}
where the ratio is determined by using a continued-fraction expansion of the golden mean 
to emulate incommensurability.
Throughout the present study we set the values of the model parameters as
\begin{eqnarray}
N_a&=&144, N_b=233, c_a=1.618\cdots, c_b=1, m_a=m_b=1,\nonumber\\
K_a&=&1, \gamma_a=\gamma_b=1.
\end{eqnarray}
We control the elasticity of chains by changing the spring constants $K_s$ and $K_b$, 
and the strength of interchain interaction by $K_I$.

In numerical simulations 
we mainly employ an initial atomic configuration where 
atoms are located at regular sites periodically. 
During the RK steps the chains are relaxed to stable states 
and finally reach there. 
In Section IV-A-5 we will refer to the effect of the initial atomic configuration 
in connection with pinning behavior. 

We employ the following numerical criterion and methods in the calculations of several quantities.
If $\dot{u}_i$ and $\dot{v}_i$ calculated during the RK steps
satisfy a velocity condition, 
$\sqrt{(\sum^{N_a}_i(\dot{u}_i)^2 + \sum^{N_b}_i(\dot{v}_i)^2 )/(N_a+N_b)}< 10^{-10}$, 
the RK calculation is stopped, and 
the state obtained then is considered to be static. 
The phonon frequency is calculated using a dynamical matrix for this stationary state.
The maximum static frictional force is evaluated as the critical force 
above which the velocity condition is not satisfied.
Using the criterion, the difference between pinned and sliding states is distinguishable.
The frictional force is calculated according to Eq. (4) or the method used in Ref.\cite{matsu1}.
In a sliding state, after the system reaches a steady state, 
the temporal average is performed in calculating the kinetic frictional force
over a time period much longer than a time during which 
the center of gravity of the system moves by the system length, $N_ac_a(=N_bc_b)$.

\section{Results}
\subsection{Pinned states}
\subsubsection{Lowest phonon frequency}

In this section we investigate the pinning effect of the two-chain model 
in the absence of external force.
To investigate the feature of pinned states, we first calculate 
the lowest-phonon frequency, which is a significant quantity because 
finite lowest-phonon frequency, i.e., the phonon gap, means the presence of pinning 
and its square is proportional to the restoring force due to pinning effects. 

For the comparison with the two-chain model, 
we first show the squared lowest phonon frequency ($\omega^2_{\rm lpf}$) 
of the FK model described by Eqs. (1) and (5) with $K_I=1$ in Fig. 1.
In this case the elastic parameter is $K_a$ only, and there exists its critical value.
Below the critical $K_a$, $\omega^2_{\rm lpf}$ becomes finite, corresponding to 
the appearance of the breaking of analyticity state due to the Aubry transition.
Above the critical $K_a$, $\omega^2_{\rm lpf}$ vanishes.
The change of $\omega^2_{\rm lpf}$ is continuous at the critical point.
Such behavior of $\omega^2_{\rm lpf}$ for the FK model was reported in Ref. \cite{peyrard}.
In the present study on the two-chain model 
we have another elastic parameters in the lower chain. 
Hence, we focus on the effect of the elastic relaxation of the lower chain, and then
$K_s$ and $K_b$ are chosen as variable elastic parameters.

In Fig. 2(a) we show $\omega^2_{\rm lpf}$ 
as a function of an elastic constant $K_b$ 
in the case of strong interchain interaction $K_I=1$, and $K_s=1$.
The strength of the interchain interaction ($K_I=1$) is chosen to be greater than 
the critical value of the Aubry transition for the FK model, 
$K_{I,{\rm FK}}^{\rm critical}\approx0.33$ for $K_a=1$. 
Because of strong interchain interaction, the phonon gap is 
finite and almost constant in the region of 
large value of $K_b$ ($2.4<K_b$).
However, steep valleys appear in the range $0.81<K_b<2.4$. 
The amount of $\omega^2_{\rm lpf}$ changes there more than two orders of magnitude. 
It is considered that $\omega^2_{\rm lpf}$ vanishes at each valley. 
The finite values of $\omega^2_{\rm lpf}$ at the bottoms of the valleys will be 
due to the numerical accuracy of the present calculation and 
the finite magnitude of changing in $K_b$. 
When $K_b$ becomes smaller than a certain value ($\approx 0.81$), 
$\omega^2_{\rm lpf}$ 
increases sharply and then becomes almost constant.
The vanishing phonon gap at each valley seems to indicate a sort of phase transition.

When the interchain interaction $K_I$ is somewhat weakened, but its strength is still greater 
than $K_{I,{\rm FK}}^{\rm critical}$, 
more drastic behavior of phonon gaps is observed.
Figure 2(b) shows $\omega^2_{\rm lpf}$ obtained for $K_I=0.45$ and $K_s=1$. 
Finite phonon gaps exist both in small and large $K_b$ regimes, 
and steep valleys appear in the intermediate regime, $0.2<K_b<0.6$. 
Such behavior is quite similar to that for $K_I=1$. 
In a wide regime, $0.6<K_b<1.4$; however, 
the phonon gap vanishes completely within a numerical accuracy. 
This regime is quite distinct from other regimes with narrow valleys 
seen in Fig. 2(a) 
as discussed in the next subsection.
The anomalous behavior of the phonon gap disappears 
when a large value of $K_s$ 
is chosen as in Fig. 2(c), where $K_s=10$ ($K_I=1$). 
It is obvious that the behavior of 
the phonon gap depends on the 
elastic parameter $K_s$ as well as $K_b$.
In Fig. 2(a) the behavior of the phonon gap looks self-similar. 
If this is the case, the phonon gap reveals more complicated behavior 
against a smaller change in $K_b$ in the intermediate $K_b$ regime.

The discontinuous and complicated behavior of $\omega^2_{\rm lpf}$ 
against the change in elastic parameters 
for the two-chain model is obviously quite different from that for the FK model shown in Fig. 1.
We confirmed that the magnitude of the phonon gap at all $K_b$'s 
chosen in Figs. 2(a)-(c) is entirely insensitive to the enlargement of 
the system size 
determined by using a continued-fraction expansion of the golden mean in Eq. (7).

\subsubsection{Maximum static frictional force}

Next we calculate the maximum static frictional force numerically 
by applying the external force to the pinned states. 
Figures 3 (a) and (b) show the maximum static frictional force calculated 
for pinned states shown in Figs. 2(a) and (b), respectively.
The maximum static frictional force also shows anomalous behavior, 
which obviously reflects the phonon gap structures in Figs. 2(a) and (b).
In Fig.3 (a), where the values of the parameters, $K_I=1$ and $K_s=1$, 
are the same with those in Fig. 2(a), 
the maximum static frictional force shows multivalley structures.
The magnitude of the maximum static frictional force 
is finite in the whole $K_b$ regime. 
We note that the values of $K_b$ at the local minima and 
maxima of the maximum static frictional force 
do not correspond exactly to those of $\omega^2_{\rm lpf}$ shown in Fig. 2(a). 
This may be considered to be the effect of the external force, 
by which the pinned lattice structures are distorted and 
the depinning threshold force would be affected slightly. 
On the other hand,
as seen in Fig. 3(b) for $K_I=0.45$ and $K_s=1$, 
the maximum static frictional force vanishes completely in the characteristic 
$K_b$ regime where the completely vanishing phonon gap is 
observed in Fig. 2(b). 
In the case that $K_I=1$ and $K_s=10$, it is confirmed that 
the maximum static frictional force 
as well as the phonon gap in Fig.2 (c) does not show any anomalous behavior 
and changes smoothly with $K_b$.
%

\subsubsection{Hull functions and lattice structures}

To investigate further the above anomalous pinning behavior observed for the phonon gap and 
the maximum static frictional force, 
we analyze the lattice structures of the pinned states by 
examining hull functions. 
Although a hull function has been employed to analyze the breaking of analyticity state 
for the FK model \cite{aubry1,peyrard}, 
it has been reported that hull functions defined in two chains are also useful to analyze 
the lattice structures both in pinned and sliding states for the two-chain model \cite{kawa1}. 
The hull functions for the two-chain model are defined as 
\begin{eqnarray}
u_i &=& i c_a +\alpha + h_a(i c_a + \alpha),\\
v_i &=& i c_b + \beta + h_b(i c_b + \beta),
\end{eqnarray}
where $h_a$ and $h_b$ are the hull functions in the upper and lower chains, respectively, 
and $\alpha$ and $\beta$ are constant phases.
The periodicities of the hull functions are expressed as 
\begin{eqnarray}
h_a(x) = h_a(x + c_b), \:\: h_b(x) = h_b(x + c_a).
\end{eqnarray}
When the chains are not deformed and hence the atoms are arrayed periodically, 
the interchain interaction potential 
is sinusoidal as mentioned in Eqs.(5) and (6), 
and then the position of the potential maximum in one period is located at 
the half of the period of the hull function, 
$x = c_b/2$ ($ c_a/2$) for $h_a(x)$ ($h_b(x)$) in our choice, $\alpha=\beta=0$.

For the convenience of later discussions, 
we briefly summarize here some features of the hull function for the FK model.
When the strength of the interchain interaction is less than the critical value of 
the Aubry transition, the hull function is smooth and continuous.
Above the critical point, however, the breaking of analyticity due to the Aubry transition 
occurs, and then the hull function changes its form 
from continuous to discrete and shows a complicated structure with many gaps. 
Among the gaps the largest one is located at the half of the period of the hull function.
The continuous form means continuous spatial atomic distribution in the underlying potential, 
and then no gap exists in the phonon excitation. 
On the other hand, the discrete one corresponds to a pinned state, 
which is accompanied by a finite gap in the phonon excitation. 
The spatial atomic distribution is vanishing at the maxima of the potential 
and the atoms are confined nearby the minima of the potential. 
Then the hull function shows the largest central gap, which 
characterizes the breaking of analyticity state for the Aubry transition. 
\\

Now we consider the case of the two-chain model. 
Figures 4(I)-(VII) show the hull functions $h_a$ and $h_b$ 
for several values of $K_b$ indicated by arrows (I)-(VII) in Fig. 2(a), 
where the magnitudes of the parameters, $K_I=1$ and $K_s=1$, 
are the same as those in Fig. 2(a).
The hull function $h_b$ in Fig. 4(I) ($K_b=0.1$; (I) in Fig. 2(a)) shows 
the largest central gap at x = $c_a/2 \approx 0.809$. 
This gap structure is essentially the same with 
that for the Aubry transition in the FK model
and indicates that the lower chain is in the conventional breaking 
of analyticity state due to the Aubry transition.
On the other hand, $h_a$ shows a discrete feature but does not have a central gap. 
Such a state in the upper chain where the central gap of 
the hull function is absent is not well defined in the context of 
the conventional breaking of analyticity state due to 
the Aubry transition of the FK model, 
but it is obviously a sort of breaking of analyticity states because of the presence of 
gaps of the hull function.
These gap structures of $h_a$ and $h_b$ remain even for $K_b=0.702$ (Fig. 4(II)). 
This reflects the constant $\omega^2_{\rm lpf}$ observed 
in the small $K_b$ regime ($K_b< 0.81$) in Fig. 2(a).
As $K_b$ increases further ($K_b > 0.81$), however, the central gap of $h_b$ is destroyed 
and no central gap exists both in $h_a$ and $h_b$.
Some other gaps also vanish or shrink, otherwise enlarge, 
and furthermore, new gaps appear at several positions.
Figures 4(III)-(V) show $h_a$ and $h_b$ at several values of $K_b$ 
where $\omega^2_{\rm lpf}$ shows a local maximum against 
the change in $K_b$ (see arrows (III)-(V) in Fig. 2(a)).
It should be noted here that the gap structures of $h_a$ and $h_b$ 
shown in (III)-(V) are different from each other. 
In a $K_b$ regime between two nearest-neighbor valleys of $\omega^2_{\rm lpf}$, 
both the gap structures of $h_a$ and $h_b$ are almost unchanged. 
Only when $K_b$ is changed crossing through the valley of $\omega^2_{\rm lpf}$, 
the gap structures are suddenly changed, i.e., new gaps appear.
In the large $K_b$ regime ($K_b>2.4$), as shown in Fig. 4(VI), 
the central gap appears in $h_a$, but it is absent in $h_b$, and 
the gaps of $h_b$ as a whole are highly reduced. 
This behavior indicates that the upper chain is in the conventional breaking 
of analyticity state due to the Aubry transition.
These gap structures of $h_a$ and $h_b$ retain up to infinite $K_b$ 
while the amplitude of $h_b$ as a whole shrinks as $K_b$ is increased (Figs. 4(VI) and (VII)). 
The $\omega^2_{\rm lpf}$ also does not change in this large $K_b$ regime.
%
%
To see how the change of gap structures of hull functions 
takes place at the valleys of $\omega^2_{\rm lpf}$, in Figs.5 (a) and (c), 
we show hull functions 
for two nearest-neighbor valleys (local minima) of $\omega^2_{\rm lpf}$ 
at $K_b= 1.18$ and 1.58.
Note here that the $K_b$ for Fig. 4(III) is located 
in the $K_b$ regime between these two nearest-neighbor valleys of $\omega^2_{\rm lpf}$.
For the comparison with these states at valleys, the graph of Fig. 4(III) 
is plotted again in Fig.5 (b).
In Figs.5 (a) and (c) 
several gaps of the hull functions observed in Fig. 5(b) are destroyed by 
the appearance of certain states in gaps, i.e., the formation of new gap structures 
at the $K_b$'s. 
As mentioned above the gap structures of hull functions for Fig. 5(b) are stable and 
almost unchanged in the regime of $1.18<K_b<1.58$ 
where no valley of $\omega^2_{\rm lpf}$ exists.
Similar behavior is observed around each valley of $\omega^2_{\rm lpf}$.
When $K_b$ reaches one of the critical value, the old gap structure becomes unstable 
and new states appear in gaps, which accompany the decrease of the phonon 
gap and then the pinning force.  
When $K_b$ crosses the critical value, new gap structure becomes 
stable and the phonon gap increases.    
Further change of $K_b$ moves the system to the next critical value 
and then the successive phase transition occurs.
%

In Fig. 6 we show the hull functions $h_a$ and $h_b$ 
for several values of $K_b$ in the case of weak interchain 
interaction, $K_I=0.45$, and $K_s=1$, which are the same values with those in Fig. 2(b).
In the small and large $K_b$ regimes 
((I), (II), and (VI), (VII) in Figs. 6, respectively) 
where $\omega^2_{\rm lpf}$ is almost constant, 
both the gap structures of $h_a$ and $h_b$ are unchanged, 
one of which shows the central gap.
This indicates that one of the upper and lower chains is in the conventional breaking of 
analyticity state due to the Aubry transition 
in the large and small $K_b$ regime, respectively.
In the intermediate regime (Figs. 6(III)-(V)), the gap structures as a whole are rather 
sensitive to the change in $K_b$, but 
also in this case the gap structures are almost unchanged in a small $K_b$ regime 
between two nearest-neighbor valleys of $\omega^2_{\rm lpf}$.
Thus, the behavior of $h_a$ and $h_b$ is similar to that in Fig. 4 in this regime.
However, in the $K_b$ regime $0.6<K_b<1.4$, for $K_I=0.45$, 
where $\omega^2_{\rm lpf}$ is vanishing, 
both the hull functions show a peculiar feature.
In Fig. 7 we show several typical $h_a$ and $h_b$ in this $K_b$ regime.
It is clearly observed that
both hull functions $h_a$ and $h_b$ are continuous 
and show sinusoidal forms, which are quite similar to that of 
the hull function for the FK model in the absence of the breaking of analyticity.
These correspond to states in which all atoms of both chains locate 
near its regular sites periodically 
and are weakly affected by the almost sinusoidal interchain force 
caused by atoms in the other chain.
Since the continuous hull functions mean that the atomic distribution is spatially continuous 
both in the upper and lower chains, 
every atom in the upper chain moves smoothly
when the upper chain is driven by the external force.
Therefore, there are no energy costs against the sliding motion of the upper chain. 
Hence, the maximum static frictional force vanishes as observed in Fig. 2(b).

For $K_s=10$ and $K_I=1$ (Figs. 8(I)-(IV)), 
$h_a$ does not show any remarkable changes of gap structures. 
$h_b$ gradually changes its gap structure only in a large $K_b$ regime ($K_b>10$), 
but the effect of the change is almost negligible because the amplitude of the gaps of $h_b$ 
becomes very small for such large $K_b$'s. 
Therefore all elastic effects come from the upper chain.
The behavior of the hull functions reflects the smooth change in the phonon gap in Fig. 1(c). 
In all regime of $K_b$, $h_a$ shows the largest central gap, but $h_b$ does not. 
Thus, in these pinned states the conventional breaking of analyticity due to 
the Aubry transition occurs in the upper chain for the whole range of $K_b$. 
Then the magnitude of the central gap in $h_a$ is quite larger than 
those of the gaps in $h_b$ and almost unchanged against the change in $K_b$.

It can be confirmed that 
the breaking of analyticity states exist in the present two-chain model, 
but they are rather complicated and 
different from the conventional one due to the Aubry transition of the FK model. 
In the small (large) $K_b$ regime, however, the gap structure of $h_b$ ($h_a$) is the same as 
the conventional one observed 
in the breaking of analyticity state in the FK model, 
while that of $h_a$ ($h_b$) is different.
That is, in the two-chain model, it is considered that 
whether the conventional breaking of analyticity state 
characterized by the largest central gap 
exists in the upper or lower chain depends on the elasticity of the two chains.
When the lower (upper) chain is highly stiffer than the upper (lower) one, i.e., 
$K_b/K_a$ or $K_s/K_a \gg 1$ ($\ll 1$), 
the atoms in the upper (lower) chain tend to relax into potential minima 
created by the atoms in the lower (upper) chain, 
and then the conventional breaking of analyticity state appears in the upper (lower) chain. 
In the intermediate $K_b$ regime, however, quite different states from 
the conventional breaking of analyticity state observed for the FK model appear.
We will again discuss this point later by calculating the energy quantities 
of the system.

It should be noted here that all the hull functions shown in Figs. 4-8 do not contain 
irregular points that break a rotational symmetry by $\pi$ of hull functions. 
This fact means that 
atomic configurations obtained above are not disordered, 
but they exactly reflect the discreteness of hull functions in an incommensurate system. 
This feature of the pinned atomic configuration is the same as that 
for the FK model \cite{peyrard}.
%

It is helpful here to observe the change of the pinned lattice structures in real space. 
Figure 9 shows the local lattice structure in the pinned state for $K_I=1$ and $K_s=1$, 
which are the same values with those in Figs. 2(a), 3(a) and 4.
Here the atomic displacements from the regular periodic sites in the two chains 
$\delta u_i$ and $\delta v_i$ are plotted 
in Figs. 9(a) and (b), respectively for $K_b$'s indicated by arrows (II)-(VI) 
in Fig. 2(a).
In the small $K_b$ regime ($K_b<0.8$), the lattice structure is essentially unchanged 
(Figs. 9(a)-(1) and (b)-(1)).
In the intermediate regime ($0.8<K_b<2.4$), however, 
both $\delta u_i$ and $\delta v_i$ are very sensitive to the change in $K_b$. 
Figures (2) and (3) in both of Figs. 9(a) and (b) correspond to atomic displacements  
at local maxima of the phonon gap $\omega^2_{\rm lpf}$ in Fig. 2(a).
It is obvious that the spatial modulation patterns of $\delta u_i$ and $\delta v_i$ 
show quasiperiodicity approximately for each $K_b$, 
but the spatial patterns are different for different values of $K_b$.
When $K_b$ increases further ($K_b > 2.5$), the reconstruction does not occur any more 
and the lattice structures as observed 
in Figs. 9(a)-(5) and (b)-(5) retain up to infinite $K_b$, 
but the atomic displacement in the lower chain 
$\delta v_i$ as a whole decreases its magnitude more and more.
Similar changes of local lattice structures are observed also 
in the intermediate $K_b$ regime for $K_I=0.45$ and $K_s=1$. 
Note here that if the behavior of the phonon gap has a self-similar nature as noticed 
in Fig. 2(a), then infinite sorts of local lattice structures would exist 
in the intermediate $K_b$ regime.
%
\subsubsection{Analysis of energy and discussion on the pinning mechanism}

We discuss further the anomalous pinning behavior by 
examining the energy of the system. 
Here it is useful to define the 
following energy quantities.
Elastic energy in the upper chain $E_{\rm ela-a}$ and that in the lower chain $E_{\rm ela-b}$ 
are given by 
\begin{eqnarray}
E_{\rm ela-a} &=& \frac{K_a}{2}\sum_i^{N_a} ( u_{i+1} - u_i - c_a )^2, \\
E_{\rm ela-b} &=& \frac{K_b}{2}\sum_i^{N_b} ( v_{i+1} - v_i - c_b )^2.
\end{eqnarray}
Elastic  energy between the lower chain and the substrate $E_{\rm ela-s}$ is given by,
\begin{eqnarray}
E_{\rm ela-s} &=& \frac{K_s}{2}\sum_i^{N_b} (v_i - i c_b)^2.
\end{eqnarray}
Total elastic energy $E_{\rm ela-total}$ is the sum of 
\begin{eqnarray}
E_{\rm ela-total} &=& E_{\rm ela-a}+E_{\rm ela-b}+E_{\rm ela-s}.
\end{eqnarray}
Interchain interaction energy $E_{\rm int}$ is expressed as 
\begin{eqnarray}
E_{\rm int} &=& \frac{1}{2}\sum_i^{N_a}\sum_j^{N_b} \left[ U_I ( u_i - v_j )
- U_I ( ic_a - jc_b ) \right],
\end{eqnarray}
where the contribution in the case of 
periodic rigid atomic configurations is subtracted. 
To evaluate Eqs.(12)-(16) we used the atomic configuration obtained in the calculation 
of $\omega^2_{\rm lpf}$.
Figures 10 (a)-(d) show these energy quantities plotted 
against $K_b$ for $K_I=1$ and $K_s=1$,
which are the same values with those in Figs. 2(a), 3(a), 4, 5 and 9.
When $K_b$ is very small ($K_b\ll 1$), the absolute value of the interchain interaction energy 
$|E_{\rm int}|$ is much greater than $E_{\rm ela-a}$, but not so much greater than $E_{\rm ela-s}$.
The upper chain deforms little, but the lower chain deforms so large 
to gain the interchain interaction.
In fact the hull function of the lower chain $h_b$ 
shows the central largest gap, but $h_a$ does not and its magnitude is small. 
These behaviors indicate that the upper chain has an almost periodic structure, 
but the lower chain adjusts its local period to that of the upper chain 
to gain the interchain interaction, that is, the lower chain forms a kind of 
discommensurate structure or the soliton lattice.
When $K_b$ is increased, the deformation of the lower chain and 
$E_{\rm ela-s}$ decrease.
On the other hand, $E_{\rm ela-b}$ increases because its coupling constant, itself, increases.
When $|E_{\rm int}|$ becomes comparable to the total elastic energy, 
the lattice structures of both chains in the small $K_b$ regime becomes unstable 
and another structure appears.
The presence of kinks for all energy curves in Fig. 10 
indicates that the structural change is a phase transition of the first order.
The difference in the lattice structures
between phases is observable directly in Figs. 9(a) and (b).
The disappearance of the central gap of the hull function $h_b$ in Fig. 4 
is attributed to the structural phase transition. 
Both chains form complicated discommensurate structures 
where both chains deform to make the local commensurate structure in order to 
gain the interchain interaction.
The occurrence of the structural phase transition leads to the sudden decrease 
(or vanishing) of the phonon gap at $K_b\approx 0.81$ as observed in Fig. 2(a). 
In the intermediate $K_b$ regime ($0.81<K_b<2.4$), 
many, or an infinite number of lattice structures appear there, 
each of which corresponds to a different discommensurate structure. 
When $K_b$ is increased small or infinitesimally, 
lattice structures change to another one by a structural phase transition. 
Such phase transitions occur successively in this regime against the change in $K_b$. 
As the lower chain becomes stiffer, the atomic displacement in the lower chain 
$\delta v_i$ is suppressed, 
and then $E_{\rm ela-b}$ and $E_{\rm ela-s}$ decrease.
Because in this $K_b$ regime a slight change in $K_b$ causes a structural phase transition, 
the phonon gap and the hull functions also change 
their structures correspondingly.
It should be noted that the absence of the central gap in both hull functions characterizes 
the lattice structures in this intermediate $K_b$ regime,
where the successive structural phase transition occurs (Figs. 4(III)-(V)). 
This indicates complicated discommensurate structures of both chains.
In the large $K_b$ regime ($K_b >2.4$), 
the structural phase transition does not occur and a quite stable 
lattice structure
appears in each chain, which is characterized by the largest central gap of $h_a$.
These behaviors  indicate that the lower chain forms an almost periodic structure, 
but the upper chain forms a discommensurate structure.
In this regime 
the major contribution to $E_{\rm ela-total}$ comes from $E_{\rm ela-a}$
because the atomic displacement $\delta v_i$ is suppressed and $\delta u_i$'s 
are fixed into a locally commensurate configuration.
Therefore, $E_{\rm ela-a}$ and $E_{\rm int}$ are almost constant.
Consequently, almost the same gap structures of the hull functions as those in Fig. 4(IV)
and the same atomic configurations as those in (5) in Figs. 9(a) and 9(b)
hold up to infinite $K_b$. 
We note here that these features, such as 
the kinks for energy and the vanishing of the phonon gap, 
which mean a first-order phase transition, are never expected in the case of 
the FK model because the Aubry transition observed in the FK model is 
a higher-order phase transition.

In Fig. 11 we show the energy quantities for the parameters $K_I=0.45$ and $K_s=1$ 
employed in Figs. 2(b), 3(b), 6, and 7. 
Also, in this case 
kinks are observed for all energy curves.
Within the $K_b$ regime ($0.6<K_b<1.4$) 
where the phonon gap and maximum static frictional force are vanishing completely, 
any anomaly such as a kink is not observed. 
It is considered that one particular lattice structure of both chains appears in this $K_b$ regime

The energy quantities obtained for $K_s=10$ and $K_I=1$ are plotted in Figs. 12 (a)-(d). 
In this case, because of the large value of $K_s$, 
the lower chain is much stiffer than the upper chain.
Then all energy quantities change smoothly with $K_b$. 
This means that 
these pinned states in all $K_b$ regimes are understood essentially 
in the context of the conventional breaking of analyticity state due to 
the Aubry transition for the FK model. 
Namely, the feature of pinned states for $K_s=10$ and $K_I=1$ are the same as 
those in the large $K_b$ regime shown in Fig. 10, where $K_s=1$ and $K_I=1$.

For simplicity, 
we have changed the elasticity of the lower chain, 
while the elasticity of the upper chain is fixed (i.e., $K_a=1$).
In general, it is considered that 
the relevant parameters to the pinning behavior of the present model 
are normalized ones: $K_b/K_a$, $K_s/K_a$, and $K_I/K_a$.

\subsubsection{Hysteretic behavior of phonon gap }

We have discussed the structural phase transitions of the first order.
Then we expect hysteretic behaviors around the critical points.
It is to be noted here that in all the results shown 
in Figs. 1-12, the initial configurations of atoms for each value of $K_b$ 
are regular and periodic.
In Fig. 13 we show phonon gaps for the same parameters
($K_I=K_s=1$) as those in Fig.2 (a) when $K_b$ is swept, that is, 
the initial atomic configuration for a certain value of $K_b$ 
is the final stable configuration for its previous value.
The solid lines with circles indicate the results obtained for 
the sweep with (a) increasing and (b) decreasing value of $K_b$.
For comparison, the results in Fig. 2(a) are also plotted by the dashed lines with squares.
By comparison, between these two results, the hysteretic behavior of the phonon gap 
is obviously observed in the intermediate $K_b$ regime.
Such hysteretic behavior, which appears suddenly only in the intermediate $K_b$ regime, is 
another evidence of the first-order phase transition.

We have considered only three initial conditions 
corresponding to a periodic atomic configuration (Figs. 2-12) and 
two history-dependent configurations (up and downward sweeps of $K_b$) (Fig. 13). 
The pinning behavior for other initial conditions has not been 
investigated exhaustively 
because a great variety of configurations may be considered.
We have found here that the hysteresis curves of phonon gaps in Fig. 13 are reproducible 
well when the starting value of $K_b$ for sweeps is set in the small or large $K_b$ 
regime.
Furthermore, even when the sweep of $K_b$ is started from the intermediate regime, 
we have observed a tendency that phonon gaps change staying on 
one of the curves shown in Fig. 13 corresponding to the increase or decrease 
in the value of $K_b$.

\subsection{Sliding states}

We here investigate the kinetic frictional force of the sliding state 
started from the pinned state.
Figure 14 shows the kinetic frictional force as a function of the sliding velocity, 
where the interchain interaction is chosen as $K_I=1$.
The line with circles shows the result in the case that $K_b=2$ and $K_s=1$, 
which correspond to the valley of the maximum static frictional force in Fig. 3(a). 
For comparison, the result for a large $K_b$ limit ($K_b=\infty$), 
where the present model corresponds to the FK model, 
is plotted in the same figure. 
The frictional force calculated by the perturbation theory \cite{soko1,kawa1} 
in the case that $K_b=2$ and $K_s=1$ 
is also plotted in this figure.
It is found that the perturbation theory explains well the numerical results 
when the interchain interaction is so weak that the maximum static frictional force 
is vanishing \cite{kawa1}.
For $K_b=2$ and $K_s=1$, the velocity-strengthening and velocity-weakening features 
of the kinetic frictional force are recovered and well explained by the perturbation theory.
In the case of the FK model ($K_b=\infty$), however, 
the velocity-strengthening feature in the low-velocity regime 
is destroyed by the large maximum static frictional force and 
the discrepancy between the simulation and the perturbation theory is obvious.
Therefore, it is considered that the effect of the anomalous pinning in stationary states affects 
the sliding velocity dependence of the kinetic frictional force especially in a low-velocity regime.

Note here that 
no hysteretic behavior is observed in the velocity dependence of 
the kinetic frictional force between increasing and decreasing processes of the driving force 
when a periodic atomic configuration is employed as the initial state of no external force. 
%

\section{Summary}

We have investigated frictional phenomena of an incommensurate two-chain model.
By controlling the elasticity of chains, 
we have found anomalous pinning behavior, which is accompanied by a successive phase transition.
The pinned states show complicated behavior against the change in elastic parameters 
and differ apparently from 
the conventional breaking of analyticity state due to the Aubry transition of the FK model.
Under certain conditions on elasticity and interchain interaction, we have confirmed that 
the pinning force and the maximum static frictional force 
are anomalously reduced or vanishing.
Such anomalous pinning behavior is quite sensitive to the elasticity of chains, 
and it is never expected for the FK model 
because the anomalous pinning occurs in a characteristic regime 
where the elasticity of both chains becomes important.

We have also found that the anomalous pinning effect affects the kinetic frictional force 
significantly. 
Also, in such a pinned state the maximum static frictional force is given in 
the vanishing velocity limit of the kinetic frictional force. 
In the case that the pinning force is weakened anomalously, 
the velocity dependence of the kinetic frictional force 
shows the velocity-strengthening and velocity-weakening features clearly, 
which are very close to those of the kinetic frictional force expected 
in the absence of the breaking of analyticity state.
In the present paper the overdamped sliding dynamics has been investigated 
in connection with the kinetic frictional force. 
For underdamped dynamics, however, the kinetic frictional force 
may show much more complicated velocity dependence, compared with the overdamped case 
\cite{kawa4}. 
It will be discussed in other reports \cite{kawa4,elmer2}.

The detailed study of the overall phase diagram in the $K_b-K_I$ plane will be 
reported elsewhere.
It might be interesting to investigate three-dimensional systems with relaxational interfaces. 
The breaking of analyticity is not restricted in one-dimensional systems and 
is also possible in higher dimensions as noticed in Refs. \cite{shinjo,hira}.
In practice, the relationship between the pinning effect and the elasticity of lattices 
in higher dimensions are an intriguing subject in tribology.
We may expect more complicated pinning behavior than 
that for the present one-dimensional model \cite{kawa3}. 

\acknowledgments

This work was financially supported by 
Grants-in-Aid for Scientific Research of the Ministry of Education, 
Science, Sports and Culture. 
The computation in this work was done using the facilities of the Supercomputer Center, 
Institute for Solid State Physics, University of Tokyo. 


\vfill\eject
%
%

\noindent
Fig. 1. \\
Squared lowest phonon frequency for the FK model ($K_a=K_I=1$ and $K_b=K_s=\infty$). 
\\

\noindent
Fig. 2. \\
Squared lowest phonon frequency vs $K_b$. 
(a) $K_I=1$ and $K_s=1$, (b) $K_I=0.45$ and $K_s=1$, and (c) $K_I=1$ and $K_s=10$. 
$K_a$ is fixed at unity. 
We will refer to insetted numbered arrows later in Figs. 5, 6 and 8.
\\

\noindent
Fig. 3.\\
Maximum static frictional force vs $K_b$. 
(a) $K_I=1$ and $K_s=1$, (b) $K_I=0.45$ and $K_s=1$.
$K_a$ is fixed at unity.
\\

\noindent
Fig. 4 \\
Hull functions for $K_I =1$, $K_a=1$, and $K_s=1$. 
The graphs in the left (right) row are the hull
functions for the upper (lower) chain. 
The values of $K_b$, 
(I) $0.1$, (II) $0.702$, (III) $1.3$, 
(IV) $1.83$, (V) $2.11$, (VI) $4.09$, and (VII) $8.14$, 
correspond to arrows indicated in Fig. 2(a).
\\

\noindent
Fig. 5 \\
Hull functions for $K_I =1$, $K_a=1$, and $K_s=1$. 
The values of $K_b$ are
(a) $1.18$, (b) $1.3$, and (c) $1.58$, where 
(a) and (c) correspond to local minima of $\omega^2_{\rm lpf}$ 
and (b) corresponds to a local maximum of $\omega^2_{\rm lpf}$ 
indicated by the arrow (III) in Fig. 2(a).
\\

\noindent
Fig. 6 \\
Hull functions for $K_I =0.45$, $K_a=1$, and $K_s=1$. 
The values of $K_b$, 
(I) $0.01$, (II) $0.15$, (III) $0.27$, 
(IV) $0.305$, (V) $0.43$, (VI) $3.21$, and (VII) $9.71$, 
correspond to arrows indicated in Fig. 2(b).
\\

\noindent
Fig. 7 \\
Hull functions for $K_I =0.45$, $K_a=1$, and $K_s=1$. 
The values of $K_b$ are 
(a) $0.8$, (b) $0.9$, and (c) $1.0$.
\\

\noindent
Fig. 8 \\
Hull functions for $K_I =1$, $K_a=1$ and $K_s=10$. 
The graphs in the left (right) row are the hull
functions for the upper (lower) chain. 
The values of $K_b$, 
(I) $0.01$, (II) $1$, (III) $10$, and (IV) $100$, 
correspond to arrows indicated in Fig. 2(c).
\\

\noindent
Fig. 9 \\
Local lattice structures of the upper and lower chains 
for $K_I=1$, $K_a=1$, and $K_s=1$. 
Local atomic displacements $\delta u_i$ and $\delta v_j$ 
are plotted against periodic regular sites $ic_a$ and $jc_b$ 
in (a) and (b), respectively.
The values of $K_b$ are 
(1) $0.702$, (2) $1.3$, (3) $1.83$, (4) $2.11$, and (5) $4.09$, 
which correspond to arrows (II), (III), (IV), (V), and (VI) 
in Fig. 2(a), respectively.
\\

\noindent
Fig. 10 \\
Energy quantities vs $K_b$ for $K_I=1$, $K_a=1$, and $K_s=1$; 
(a) $E_{\rm ela-a}$, (b) $E_{\rm ela-b}$, (c) $E_{\rm ela-s}$, and (d) $E_{\rm int}$.
\\

\noindent
Fig. 11 \\
Energy quantities vs $K_b$ for $K_I=0.45$, $K_a=1$, and $K_s=1$; 
(a) $E_{\rm ela-a}$, (b) $E_{\rm ela-b}$, (c) $E_{\rm ela-s}$, and (d) $E_{\rm int}$.
\\

\noindent
Fig. 12 \\
Energy quantities vs $K_b$ for $K_I=1$, $K_a=1$, and $K_s=10$; 
(a) $E_{\rm ela-a}$, (b) $E_{\rm ela-b}$, (c) $E_{\rm ela-s}$, and (d) $E_{\rm int}$.
\\

\noindent
Fig. 13 \\
Squared lowest phonon frequency vs $K_b$ for $K_I=K_s=1$. 
The solid lines with circles indicate the results obtained for 
the sweep with (a) increasing and (b) decreasing values of $K_b$.
For comparison, the results in Fig. 2(a) are also plotted by the dashed lines with squares.
\\

\noindent
Fig. 14 \\
Kinetic frictional force vs sliding velocity. 
Triangles denote numerical results obtained in the case that 
$K_b \rightarrow \infty$, $K_a=1$, and $K_I=1$, i.e., 
in a case of the FK model. 
Circles and dotted line denote a numerical result and 
an analytic result calculated with perturbation theory, respectively, 
for $K_I=1$, $K_a=1$, $K_b=2$, and $K_s=1$.


\begin{references}
\bibitem{elmer2} For the FK and FK-Tomlinson models, 
the underdamped sliding dynamics was investigated by Strunz, Weiss, and Elmer in Ref.\cite{elmer}.
\end{references}
\end{document}